\documentclass[a4paper]{jpconf}
\usepackage{graphicx}
\begin{document}
\title{A Validation Framework for the Long Term Preservation of High Energy Physics Data}

\author{Dmitri Ozerov, David M. South}

\address{Deutsches Elektronen Synchrotron, Notkestrasse 85, 22607 Hamburg, Germany}

\ead{dmitri.ozerov@desy.de, david.south@desy.de}

\begin{abstract} The study group on data preservation in high energy physics, DPHEP, is moving to a new
collaboration structure, which will focus on the implementation of preservation projects, such as those
described in the group's large scale report published in 2012.
One such project is the development of a validation framework,  which checks the compatibility of
evolving computing environments and technologies with the experiments software for as long as possible,
with the aim of substantially extending the lifetime of the analysis software, and hence of the usability
of the data.
The framework is designed to automatically test and validate the software and data of an experiment against
changes and upgrades to the computing environment, as well as changes to the experiment software itself.
Technically, this is realised using a framework capable of hosting a number of
virtual machine images, built with different configurations of operating systems and the relevant software,
including any necessary external dependencies.

\end{abstract}

\section{Introduction: Data preservation in high energy physics}

The problem of data persistence and preservation is not new, but is becoming more prominent with the 
advent of so called big data, in particular within the applied sciences. However, until recently high energy physics
(HEP) had little or no tradition or clear model of long-term preservation of data in a meaningful and useful way,
and the data from the majority of older experiments have simply been lost. Attempts to preserve previous data
sets have in general not been a planned initiative by the original collaboration but a  push by knowledgeable
people, usually at a later date. This is despite several clear scenarios where preservation of HEP data is beneficial
for a number of reasons including: to allow the re-analysis of data taken at a unique centre of mass energy and/or
with unique initial state particles, especially if new theoretical predictions or analysis techniques become available;
to aid the combination of data sets between similar experiments; for verification of new
phenomena found by another HEP experiment; and to allow the use of real HEP data in scientific training,
education and outreach. 
 
The start of the 21st century saw the end of operation of several particle colliders including LEP ($e^{+}e^{-}$, where
data taking ended in 2000), HERA ($e^{\pm}p$, 2007), PEP-II ($e^{+}e^{-}$, 2008), KEKB ($e^{+}e^{-}$, 2010) and the
Tevatron ($p\bar{p}$, 2011), providing unique data sets in terms of initial state particles or centre of mass energy 
or both. As the experiments at each of these colliders continued to publish their final results and conclude their
core physics programmes, the question of what should be done with the data naturally presented itself. Inspired
by a lack of concrete solutions or guidelines to the problem of data preservation in HEP, an international study group
on data preservation and long-term analysis in high-energy physics, DPHEP, was formed at the end of 2008 to address
the issue in a systematic way.

The composition of the group was initially driven by BaBar and the HERA experiments H1, ZEUS and 
HERMES, who were soon joined by Belle, BES-III and the Tevatron experiments CDF and D{\O}. The LEP experiments
are also represented in DPHEP and the LHC experiments ALICE, ATLAS, CMS and LHCb joined the study group in 2011.
The laboratories and associated computing centres at BNL, CERN, DESY, Fermilab, JLAB, KEK and SLAC are also
all members of DPHEP, in addition to several funding agencies. A series of seven workshops have taken place since
2009 and DPHEP is officially endorsed with a mandate by the International Committee for Future Accelerators, ICFA. 

The initial findings of the study group are summarised in a short interim report released in 2009 and a full 
status report was published in 2012~\cite{blueprint}. The report contains: a tour of data preservation activities
in other fields; an expanded description of the physics case; a guide to defining and establishing data
preservation principles; and updates from experiments and joint projects, as well as person-power estimates for
these and future projects. After many decades of neglect with respect to other scientific disciplines, data
preservation is now a rapidly emerging field in HEP, where DPHEP is established as the coherent multi-laboratory,
multi-experiment body to examine this issue. 

The DPHEP Study Group is now moving to a new organisational model, the DPHEP Collaboration and the 
formal signing procedure of the Collaboration Agreement has now commenced. In addition to the DPHEP Chair,
a Project Manager position has now been established, which is initially based at CERN.

\section{Preservation models}

If one thing may be learned from previous enterprises, it is that the conservation of tapes is not equivalent
to data preservation, and that not only the hardware to access the data but also the software and
environment to understand the data are the necessary and more challenging aspects. In addition to the data,
the various software, such as simulation, reconstruction and analysis software need to be considered. If the
experimental software is not available the possibility to study new observables or to incorporate new 
reconstruction algorithms, detector simulations or event generators is lost. Without a well defined and 
understood software environment the scientific potential of the data may be limited. Just as important are 
the various types of documentation, covering all facets of an experiment.
Considering this inclusive definition of HEP data, the DPHEP Collaboration has established a series of data
preservation levels, as summarised in table~\ref{tab:levels}.
The levels are organised in order of increasing benefit, which comes with increasing complexity and cost.
Each level is associated with use cases, and the preservation model adopted by an experiment should reflect
the level of analysis expected to be available in the future. 

\begin{table}[t]
  \begin{center}
    \caption{\label{tab:levels}Data preservation levels as defined by the DPHEP Collaboration.}
    \centering
    \begin{tabular}{{c}{l}{l}}
      \br
      Level & Preservation Model & Use Case\\
      \br
      1 & Provide additional documentation & Publication related info search \\
      \mr
      2 & Preserve the data in a simplified format & Outreach, simple training analyses\\
      \mr
      3 & Preserve the analysis level software and & Full scientific analyses,\\
      & data format & based on the existing reconstruction\\
      \mr
      4 & Preserve the simulation and reconstruction & Retain the full potential of the\\
      & software as well as basic level data & experimental data\\
      \br
    \end{tabular}
  \end{center}
\vspace{-0.6cm}
\end{table}

The four levels represent three different areas, representing complementary initiatives: documentation 
(level 1), outreach and simplified formats for data exchange (level 2) and technical preservation projects 
(levels 3 and 4). Whereas most collaborations involved in DPHEP pursue some form of level 1 and 2 
strategies, levels 3 and 4 are really the main focus of the data preservation effort: to maintain usable access 
to analysis level data, Monte Carlo and the analysis level software, in addition (in the case of level 4) to the 
reconstruction and simulation software. Previous data preservation experiences in HEP indicate that new 
analyses and complete re-analyses are only possible if all the necessary ingredients to retrieve, reconstruct 
and understand the data are accounted for. Only with the full flexibility provided by a level 4 preservation model
does the full potential of the data remain.

Most experiments in DPHEP plan for a level 4 preservation programme, although different approaches are 
employed concerning how this goal should be achieved.  One option, typically realised using virtualisation, 
is to freeze the current system to keep the software and environment alive as long as possible. Although 
this will provide a workable solution for the medium-term future, the operability of the software and 
correctness of the results are not guaranteed. If changes are needed it will become more difficult the longer 
software is frozen and future compatibility issues may arise, in addition to the inevitable security concerns.

The alternative approach employed at DESY is to adapt and validate the software and environment to 
future changes as and when they happen. In this way, the working version of the experimental software is 
actively migrated to more modern platforms and future-proof resources, substantially extending the lifetime 
of the software, and hence the data, whilst retaining the flexibility to apply any necessary changes. The 
success of such migrations depends on having a robust and complete set of validation tests. Virtualisation 
techniques are again used, in this case to provide a coherent platform to build and validate software and 
environments created according to the recipes provided by the experiments. 

\section{The software preservation system at DESY}

A generic validation suite, which includes automated software build tools and data validation, has been 
developed at DESY to automatically test and validate the software and data of an experiment against
changes and upgrades to the environment, as well as changes to the experiment software itself. Technically, 
this is realised using a framework capable of hosting a number of virtual machine 
images, built with different configurations of operating systems (OS) and the relevant software, including 
any necessary external dependencies.

\begin{figure}[t]
  \begin{center}
    \includegraphics[width=\textwidth]{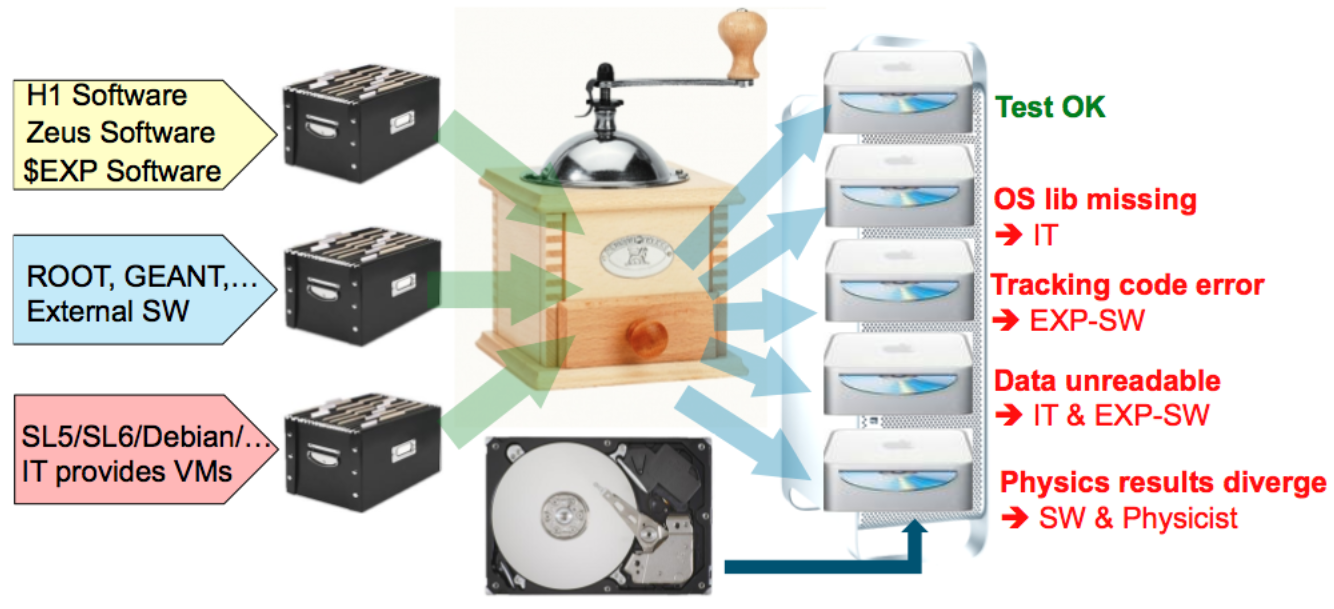}
  \end{center}
  \vspace{-0.4cm}
  \caption{An illustration of the validation system developed at DESY. Note the clear separation
  of the inputs: experiment specific software, external dependencies and operating system.}
  \vspace{-0.2cm}
 \label{fig:spsystem}
\end{figure}

\subsection{Inputs to the preservation system and work flow}

Three distinct categories are identified as separate inputs to the validation system, as illustrated in
figure~\ref{fig:spsystem}: the experiment specific software, any external software dependencies and finally
the operating system, including the compiler. The work flow of the validation framework, called the {\it sp-system},
is then as follows: 
 
\begin{enumerate}

\item In an initial, preparatory phase, the experimental software should be consolidated, the OS migrated 
to the most recent release, and any unnecessary external dependencies removed. Any remaining, 
well-defined necessary dependencies are then also incorporated. Analysis and data validation tests 
should then be defined and prepared, examining each part of the experimental software deemed 
necessary in the preservation model adopted. 
 
\item A regular build of the experimental software is done automatically according to the current 
prescription of the working environment, and the validation tests are performed. At regular intervals, 
new OS and software versions will then be integrated into the system, under the supervision of 
experts from the host IT department and experiment. 
 
\item If the validation is successful, no further action must be taken. If a test fails, any differences 
compared to the last successful test are examined and problems identified. Intervention is then 
required either by the host of the validation suite or the experiment themselves, depending on the 
nature of the reported problem. 
 
\item The final phase occurs either when no person-power is available from the experiment or IT side or 
the current system is deemed satisfactory for the long-term need or stable enough. At this point the 
last working virtual image is conserved and constitutes the last version of the experimental software 
and environment. It should be noted however, that this now frozen system is unlikely to persist in a 
useful manner much beyond this point. 
 
\end{enumerate}

The sp-system is designed for software verification, validation and migration support only. Neither the 
hardware resources nor the interface are designed for mass production or large-scale analysis. The 
framework is rather used to establish the latest working version of the computing and software 
environment and it can help to prepare a production system by supplying the successfully validated recipe 
of the latest configuration. If a production system is required, then this recipe should be deployed on a 
suitable resource at the time: an institute cluster, grid, cloud, sky, quantum computer, and so on. 
 
Within the current sp-system there are virtual machines with five different configurations: SL5/32bit with 
gcc4.1 and gcc4.4, SL5/64bit with gcc4.1 and gcc4.4, SL6/64bit with gcc4.4. In addition, the set of 
external software required by the experiments is also installed, for example the ROOT versions used by the 
experiments: 5.26, 5.28, 5.30, 5.32, and 5.34. The sp-system is designed and constructed in a such a way 
that new client machines (as a virtual machine or a normal physical machine like a batch or grid worker 
node) can easily be added. The only requirement of a new machine is to have access to the common sp- 
system storage where the tests from the experiments as well as the test results are stored, as well as the 
ability to run a cron-job on the client. 
 
\subsection{Validation tests as defined by the experiments}

In addition to the common infrastructure provided by the IT department, the development and implementation 
of the tests by the participating experiments requires significant investment, even if basic validation 
structures already exist. As a first step, the number and nature of the experimental tests is surveyed, the 
level of which reflects the DPHEP preservation level aimed at the participating collaboration.  

\begin{figure}[t]
  \begin{center}
    \includegraphics[width=0.8\textwidth]{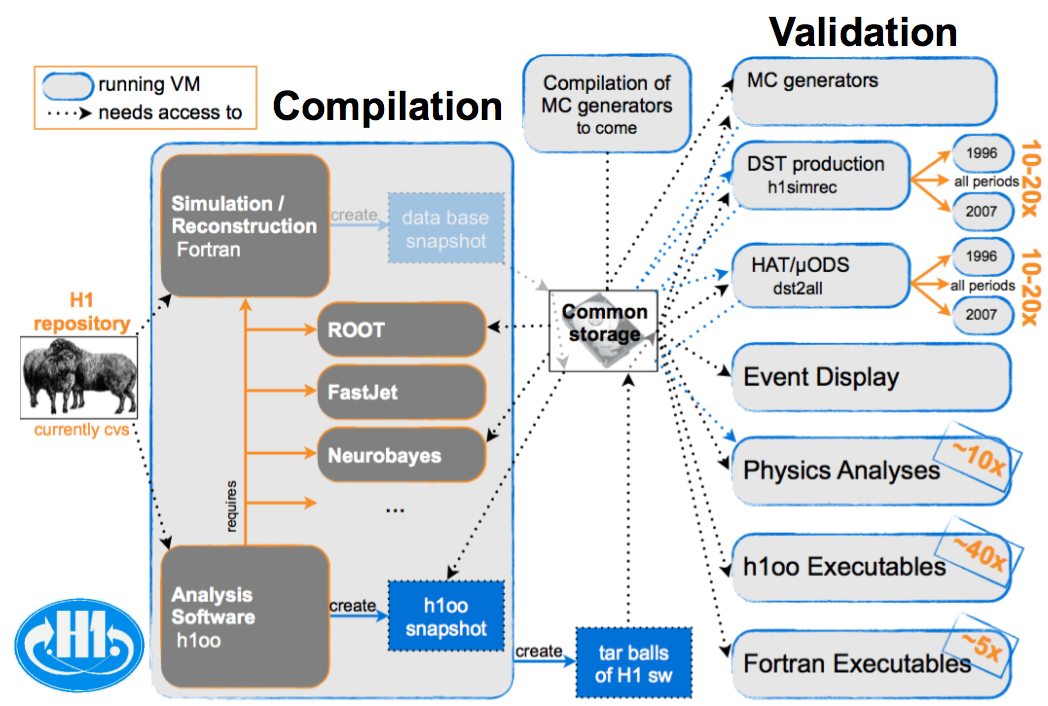}
  \end{center}
  \vspace{-0.4cm}
  \caption{An outline of the validation tests to be prepared by the H1 experiment.}
  \vspace{-0.2cm}
 \label{fig:tests}
\end{figure}

Figure~\ref{fig:tests} details the results of this evaluation by the H1 Collaboration, which is implementing a full
level 4 preservation programme, expected to comprise of up to 500 tests in total. The structure of the tests is
divided into two parts. Firstly the compilation of  approximately 100 individual H1 software packages and the
identified external dependencies is carried out, where the resulting binaries are stored as tar-balls on the common
storage within the sp-system.
Secondly, a series of validation tests is performed on the full spectrum of the H1 software, using the compiled 
software. Whereas some of these tests examine the results of stand alone executables and are run in 
parallel, many are run sequentially and form discrete parts in one of several full analysis chains: from MC 
generation and simulation, through multi-level file production and ending with a full physics analysis and 
subsequent validation of the results.

\subsection{Validation results}

Each test-job started in the sp-system is typically assigned a unique ID, and all scripts and input files used 
in the test as well as all output files are kept. This allows the validation of all versions against each other 
and ensures reproducibility of previous results. In addition to this unique ID, validation jobs may be tagged 
with a description, indicating which software versions were used, and the Unix time stamp of the execution 
to aid the bookkeeping. Script-based web pages are used to record and display available validation runs for 
a given description and indicate the status of the compilation for the individual packages or tests within 
table cells, which are linked to a corresponding output file. This file may be a simple yes/no, a text file, a 
histogram, a root file or even a link to a further page, depending on the nature of the test.

\begin{figure}[t]
  \begin{center}
    \includegraphics[width=0.8\textwidth]{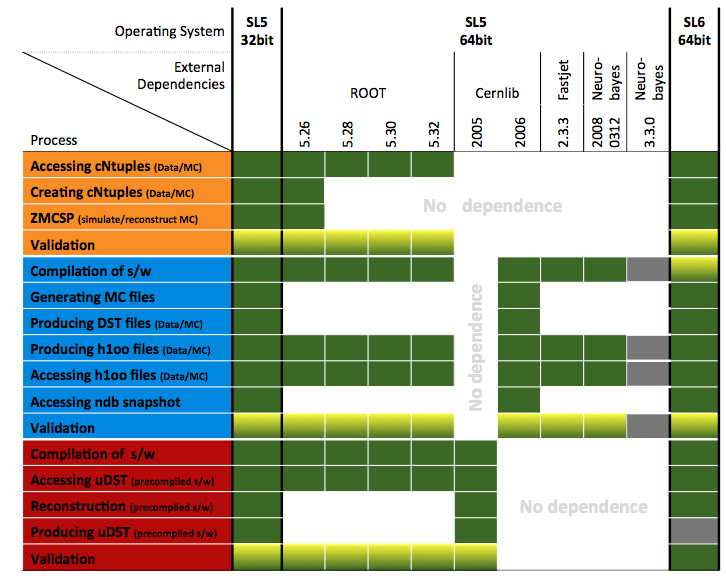}
  \end{center}
  \vspace{-0.4cm}
  \caption{A summary of the validation tests carried out by the HERA experiments within the sp-system at DESY.
    The different tests (processes) from the ZEUS (orange, top), H1 (blue, middle) and HERMES (red, bottom)
    experiments are run under different configurations of operating system and external dependencies.}
  \vspace{-0.2cm}
 \label{fig:results}
\end{figure}

In total more than 300 runs over sets of pre-defined tests have been performed within the sp-system by the 
HERA experiments. A summary of the current status of the validation tests is displayed in figure~\ref{fig:results}, 
showing a coarse breakdown for ZEUS (orange), H1 (blue) and HERMES (red) tests and the different 
dependencies. The experiments are in the process of migrating to SL6/64bit, and the tests performed so far 
using the sp-system have already identified and helped to solve several long-standing bugs. The next 
challenges include the testing of the SL7 environment and checking the compatibility of the experiments
software with ROOT 6. 

 \section{Current situation and opportunity for expansion}

A software preservation system has been developed at DESY, supporting a series of validation tests
defined by the HERA experiments, to ensure that these unique data are available for analysis for the
next 10 years or more.
The process of defining and implementing the complete set of validation tests for the whole chain of
software to be preserved is still ongoing and is expected to take another year.
From the IT side, the current version of the sp-system is very light for the user tests implemented
so far and the common storage allows communication between the sp-system and the experiment tests
using only a few shell variables.
These variables describe for example the location of the input file of the tests, the test outputs and the
external software on the client.
Using thin layers of scripts, a separation of the user part from the details of the sp-system is possible,
allowing already existing user tests to be integrated into the sp-system or tests developed within the sp-system
to be ported to other test platforms.
By design the sp-system is expandable and able to host and validate the requirements of multiple experiments,
and can be thought of as a tool to aid migration in order to detect problems and incoherence, as well as
identifying and reporting the reasons behind them.

\end{document}